\documentclass[prd,showpacs,preprintnumbers,amsmath,amssymb,floatfix]{revtex4-1}

\usepackage{graphicx}
\usepackage{dcolumn}
\usepackage{bm}
\usepackage{amssymb}
\usepackage{epsfig}
\usepackage{color}

\newcommand{\ba}{\begin{eqnarray}}
\newcommand{\ea}{\end{eqnarray}}
\newcommand{\be}{\begin{equation}}
\newcommand{\ee}{\end{equation}}
\newcommand{\bdisplay}{\begin{displaymath}}
\newcommand{\edisplay}{\end{displaymath}}

\newcommand{\eq}[1]{Eq.\,(\ref{#1})}
\newcommand{\fig}[1]{Fig.\,\ref{#1}}

\makeatletter
\def\eqnarray{\stepcounter{equation}\let\@currentlabel=\theequation
\global\@eqnswtrue
\tabskip\@centering\let\\=\@eqncr
$$\halign to \displaywidth\bgroup\hfil\global\@eqcnt\z@
  $\displaystyle\tabskip\z@{##}$&\global\@eqcnt\@ne
  \hfil$\displaystyle{{}##{}}$\hfil
  &\global\@eqcnt\tw@ $\displaystyle{##}$\hfil
  \tabskip\@centering&\llap{##}\tabskip\z@\cr}

\def\endeqnarray{\@@eqncr\egroup
      \global\advance\c@equation\m@ne$$\global\@ignoretrue}

\def\@yeqncr{\@ifnextchar [{\@xeqncr}{\@xeqncr[5pt]}}
\makeatother

\begin{document}

\title{Coulomb-nuclear interference effects in proton-proton scattering: A simple new eikonal approach.}

\author{Loyal Durand}
\email{ldurandiii@comcast.net}
\altaffiliation{Mailing address: 415 Pearl Ct., Aspen, CO 81611}
\affiliation{Department of Physics, University of Wisconsin, Madison, WI 53706}
\author{Phuoc Ha}
\email{pdha@towson.edu}
\affiliation{Department of Physics, Astronomy and Geosciences, Towson University, Towson, MD 21252}
\begin{abstract}
We present a simple new approach to the treatment of Coulomb-nuclear interference and form-factor effects in high-energy proton-proton scattering in the context of an eikonal model for the spin-averaged scattering amplitude. We show that the corrections to the nuclear and Coulomb amplitudes do not depend sensitively on the details of the eikonal amplitude and can be taken as universal, and present parametrizations for the necessary corrections. We also present a simple model for the nuclear scattering amplitude useful for data analysis at small momentum transfer which builds in the proper nuclear phase and the diffraction zeros in the real and imaginary parts of the amplitude.
\end{abstract}
%
%
\maketitle


\section{Introduction \label{sec:introduction}}

The interplay of Coulomb and nuclear interactions in the scattering of charged particles has been studied by many authors \cite[and further references therein]{BetheCoulomb,WestYennie,Cahn,Islam,KL-Coulomb,Buttimore,KopeliovichTarasov,Selyugin1, Selyugin2} with particular emphasis on the use of Coulomb-nuclear interference effects to determine the real part of the nuclear scattering amplitude in proton-proton and proton-antiproton scattering at high center-of-mass energies $W=\sqrt{s}$ and small squares of the covariant momentum transfer $q^2=\lvert t\rvert$.  The results most used in recent analyses of spin-averaged high-energy data appear to be those of Cahn \cite{Cahn} as modified by Kundr\'{a}t and Lokaji\v{c}ek  \cite{KL-Coulomb}. In their approach, the Coulomb and purely nuclear effects are separated out in a spin-independent scattering amplitude, with its components expressed in terms of convolutions involving the nuclear and Coulomb amplitudes with the effects of the  proton electromagnetic form factors included. The result is rather cumbersome to use, especially because the way in which the amplitudes are separated leaves long-range effects associated with the Coulomb amplitude in some terms. Questions have also been raised about the treatment of the form factors \cite{Petrov1,Petrov2}; see also \cite{Kaspar}, and \cite{TOTEM2016} for further references.

In the present paper, we present an analysis of the Coulomb and form-factor effects in $pp$ scattering based on an  eikonal model for the spin-averaged $pp$ scattering amplitude in the limit in which spin effects can be ignored, as discussed in the Appendix.  Possible spin effects were considered in detail by Buttimore, Gotsman, and Leader in \cite{Buttimore} using a small-$\lvert t\rvert$ or small-$q^2$ expansion.  We show in the Appendix that these effects are likely to be negligible at the energies of current interest in $pp$ scattering, $W\gtrsim 100$ GeV. 

Our eikonal approach is based on a realistic model which fits the $pp$ and $\bar{p}p$ data from 4.5 GeV to cosmic ray energies, and is consistent with the phase constraints imposed by analyticity \cite{eikonal2015}. The model allows us to calculate the Coulomb and form-factor effects in the scattering without significant approximation at any value of $q^2$ for which it holds, extending beyond the first diffraction minimum in the differential cross section. No small-$q^2$ expansion such as those used in \cite{Buttimore} and, over an extended range, in \cite{KopeliovichTarasov}, is necessary. 

Our separation of the various effects in the scattering is different from that of Cahn \cite{Cahn}, with long-range effects appearing only in a pure Coulomb scattering term with unmodified form factors, and---to high accuracy---with the remaining effects isolated in a nearly model-independent phase factor that modifies the purely nuclear term,
\be
\label{final_amp}
 f(s,q^2) \approx -\frac{2\eta}{q^2}F^2(q^2) + e^{i\Phi_{tot}(s,q^2)}f_N(s,q^2).
 \ee
As is usual in the treatment of infinite-range Coulomb scattering, an irrelevant overall phase has been absorbed. In contrast to \cite{Buttimore}, we can treat the various corrections to the spin-independent terms completely, without their expansion in $t$ and $\alpha\ln\lvert t\rvert$.

The advantage of this form of the amplitude is that the Coulomb term is real, making it clear that the Coulomb-nuclear interference depends only on the real part of the second term, that is, on the real part of $f_N(s,q^2)$ with a (small) admixture of the imaginary part dependent on the phase $\Phi_{tot}(s,q^2)$. The latter is essentially model- independent for any eikonal model consistent with the the measured $pp$ total cross sections, the forward slope parameters $B=-d\log\left(d\sigma/dq^2\right)/dq^2$, and the diffractive structure at larger $q^2$. We find that $\Phi_{tot}(s,q^2)$ is small and easily parametrized in the $q^2$ region inside the first diffraction zero.

In the following sections, we first discuss the separation of the various effects in the scattering in Sec.\ \ref{subsec:background}, and then evaluate the Coulomb scattering term in eikonal form in Sec.\ \ref{subsec:coulomb}, and the form-factor effects in Sec.\ \ref{subsec:formfactors}. We show, in particular, that the Coulomb and form-factor effects combine to high accuracy to give a combined amplitude of the form $-(2\eta/q^2)F_Q^2(q^2)e^{i\Phi_{c,FF}}$ where $\eta=\alpha/v\approx\alpha$, $F_Q(q^2)$ is the charge form factor of the proton, and $\Phi_{c,FF}$ is a known phase.

We evaluate the remaining nuclear-dependent term in Sec.\ \ref{subsec:CoulNuclear}, where we show that the effects of the Coulomb interaction and the form factors combine at small $q^2$ to simply multiply the nuclear amplitude by a phase factor $e^{-\Delta\Phi(s,q^2)}$. The Coulomb and modified nuclear amplitudes can then be combined in the form in \eq{final_amp}, with an overall phase absorbed. We also present an accurate---and essentially model-independent---parametrization of the phase $\Phi_{tot}$ in \eq{final_amp} valid from energies below 100 GeV to 20 TeV or above. The result is a very simple implementation of the Coulomb and form-factor corrections in the complete scattering amplitude.

In Sec.\ \ref{sec:application}, we apply these results to analyze a model used in recent fits to Coulomb-nuclear interference at very high energies \cite{TOTEM2016,TOTEM2019} in which the phase of the nuclear amplitude is taken as constant. We show that this leads to efffective real parts of the scattering amplitude significantly larger than the actual real parts. It is very simple to improve the model at high energies by using an approximate nuclear phase that takes into account the zeros in the real and imaginary parts of $f_N(s,q^2)$ which lie within or close to the region in $q$ used in the fits, and give parametrizations of the locations of those zeros in our eikonal model valid from $\sim500$ GeV to above 20 TeV.

Finally, for completeness, we discuss the connection of the spin-averaged scattering amplitude to the complete spin-dependent elastic scattering matrix in the Appendix.


\section{Analysis \label{sec:analysis}}

\subsection{Background \label{subsec:background}}


In the absence of significant spin effects---thought to be very small at high energies---the spin-averaged differential cross section  for proton-proton scattering can be written in terms of a single  spin-independent amplitude
\be
\label{f^tot}
f(s,q^2) = i\int_0^\infty db\,b\left(1-e^{2i\left(\delta_c^{tot}(b,s)+\delta_N(b,s)\right)}\right)J_0(qb),
\ee
as sketched in the Appendix. Here $q^2=-t$ is the square of the invariant momentum transfer, $b$ is the impact parameter, $\delta_c^{tot}(b,s)$ is the full Coulomb phase shift including the effects of the finite charge structure of the proton,   $\delta_N(b,s)$ is the nuclear phase shift, and
\be
\label{delta_c^tot}
\delta_c^{tot}(b,s) = \delta_c(b,s)+\delta_c^{FF}(b,s),
\ee
where $\delta_c$ gives the phase shift for a  pure Coulomb interaction, and $\delta_c^{FF}$ accounts for the effects of the form factors at large momentum transfers or short distances. This form, with (approximately) additive phase shifts, and that in \eq{f^tot}, can be derived  in the context of potential scattering through a Glauber-type treatment \cite{Glauber} of the eikonal function.

With our normalization, the differential elastic scattering amplitude is
\be
\label{dsigma/dq^2}
\frac{d\sigma}{dq^2}(s,q^2)= \pi\lvert f(s,q^2)\rvert^2.
\ee

The finite proton charge structure appears through the charge form factors $F_Q(q^2)$ measured in electron-proton scattering \protect\footnote{ In the Glauber construction \cite{Glauber}, the Coulomb interaction $V({\bf r})$ appears as a convolution of the finite charge distributions of the protons with the Coulomb denominator, with a Glauber phase
\bdisplay
\int_{-\infty}^\infty dz\,V_c({\bf r}) = \alpha\int_{-\infty}^\infty dz\,\int d^3r_1\int d^3r_2 \,\rho(r_1)\frac{1}{\lvert {\bf r-r_1+r_2}\rvert}\rho(r_2).
\edisplay
 After introducing transverse and longitudinal coordinate ${\bf b}_i,\,z_i$, integrating over the $z$'s, and applying the Fourier-Bessel transform in the Glauber construction, the leading term in the Glauber amplitude reduces to the expected form in momentum space, $(\alpha/q^2)F^2(q^2)$, where the form factors $F(q^2)$ are Fourier transforms of the non-relativistic charge distributions $\rho$. (In the relativistic theory, $F_Q(q^2)$ can be identified as a three-dimensional Fourier transform of the time-averaged charge distribution in the Breit or brick-wall coordinate frame.\cite{VertexFuncs}) To isolate the effects of the form factors in the eikonal function, we will rewrite $V_c$ as $\alpha/r-\left[\alpha/r-V_c({\bf r})\right] $, isolating the long-range part of the Coulomb interaction in the first term. The second term, which vanishes for $r\rightarrow\infty$, gives the phase $2\delta_c^{FF}$ in \eq{delta_c^tot}, the first term, a pure Coulomb phase. The nuclear phase shift adds on separately as in \eq{f^tot}. The connection between the eikonal approach and the Born series for the scattering amplitude is discussed in \protect\cite{Glauber}.}. Only the charge form factor appears. The magnetic moment scattering with form factor $F_M(q^2)$ appears only in the spin-dependent part of the scattering amplitude  and does not contribute to Coulomb-nuclear interference in the spin-averaged cross section except through interference effects in  the average of the spin-dependent terms, thought to be very small and neglected here; see, {\em e.g.}, \cite{VertexFuncs} and \cite{Buttimore}. This is discussed further in the Appendix. The exact spin-averaged Coulomb amplitude for proton-proton scattering should therefore reduce to
 \be
\label{f_Born}
f_c^B(s,q^2) = -\frac{2\eta}{q^2}F_Q^2(q^2)
\ee
in Born approximation.

 \eq{f^tot} can be rearranged in the form
\ba
\label{f^tot2}
f(s,q^2) &=& f_c(s,q^2) + i\int_0^\infty db\, b\, e^{2i\delta_c(b,s)}\left(1-e^{2i\delta_c^{FF}(b,s)}\right)J_0(qb) \nonumber \\
\label{C+N_amp}
&& +i\int_0^\infty db\,b\,e^{2i\delta_c(b,s)+2i\delta^{FF}_c(b,s)}\left(1-e^{2i\delta_N(b,s)}\right)J_0(qb).
\ea
Here $f_c(s,q^2)$ is the  Coulomb amplitude  without form factors.
The second term in \eq{C+N_amp}, which we will label $f_c^{FF}$, accounts for the effects of the form factors on the Coulomb scattering, strongly reducing the $1/q^2$ falloff of the pure Coulomb term at large $q^2=\lvert t\rvert$. The final term $f_{N,c}$ includes the effects of the nuclear scattering as modified by the Coulomb and form factor effects. We will consider these individually in the following subsections. The pure nuclear amplitude $f_N(s,q^2)$ is just
\be
\label{f_N}
f_N(s,q^2) = i\int_0^\infty db\,b\left(1-e^{2i\delta_N(b,s)}\right)J_o(qb).
\ee
%


\subsection{Coulomb scattering \label{subsec:coulomb}}


The Coulomb phase shift depends on the parameter $\eta=z_1z_2\alpha/v$, $v=2pW\big/(W^2-2m^2)$ the velocity of either particle in the rest frame of the other, here expressed in terms of the total center-of-mass energy $W=\sqrt{s}$ and the  corresponding proton momentum $p=\sqrt{W^2/4-m^2}$; clearly $\eta\approx\alpha\ll1$ for protons at high energies. The phase shift  for the orbital angular momentum $L$ is given to leading order in $\alpha$ by \cite[Sec.\,14.6]{Newton}
\ba
\label{delta_c}
e^{2i\delta_c} &=& \frac{\Gamma(L+1+i\eta)}{\Gamma(L+1-i\eta)} = \exp{\left(2i\sum_{k=1}^L\arctan\frac{\eta}{k}\right)}  \\
&=& \exp{\left(2i\sum_{k=1}^L\left(\frac{\eta}{k}-\frac{1}{3}\frac{\eta^3}{k^3}+\cdots\right)\right)} \approx \exp{2i\left(\eta\ln{L}+\eta\gamma+O\left(\frac{\eta}{L}, \eta^3\right)\right)}
\ea
Thus, for $L=pb$ large, $b$ the impact parameter in the scattering, the Coulomb phase factor to first order in $\eta$ is
\be
\label{e^2idelta_c}
e^{2i\delta_c(b,s)} = e^{2i\eta(\ln{pb}+\gamma)} = (pb)^{2i\eta}e^{2i\eta\gamma}
\ee
where $\gamma=0.5772\ldots$ is Euler's constant.

The result for $f_c(s,t)$ follows from \eq{f^tot} for  pure Coulomb scattering, $\delta_c^{FF}=\delta_N=0$. This gives
\ba
\label{f_c_int1}
f_c(s,t) &=& i\int_0^\infty db\,b\left(1-e^{2i\delta_c(s,b)}\right)J_0(qb) \nonumber \\
 &\longrightarrow&  -ie^{2i\eta\gamma}\int_0^\infty db\,b (pb)^{2i\eta}J_0(qb),
\ea
where we have dropped the delta function in $q$ associated with the 1 in the first term in the integrand and restricted our attention to angles away from the extreme forward direction, $q> 0$. This is the usual restriction for Coulomb scattering, necessitated by the infinite range of the interaction and the associated failure of the term $e^{2i\delta_c}$ to vanish for $b\rightarrow \infty$ \cite[Sec.\,14.6]{Newton}.

To evaluate the apparently divergent integral in \eq{f_c_int1}, we note that the phase factor in \eq{delta_c} and the scattering amplitude in \eq{f_c_int1} are  analytic in $\eta$ and the phase is non-singular and nonzero for $\lvert \Im\,\eta\rvert<L+1$ or $\lvert\Im\,\eta\rvert<pb+1$. We can therefore take $\Im\,\eta\gtrsim\frac{1}{2}$  initially for $pp$ scattering ($\eta=+\alpha/v$). The integral then converges and gives  \cite[Eq.\,10.22.43]{DLMF}
\ba
\label{f_c_int2}
f_c(s,t) &=& -\frac{2\eta}{q^2}\left(\frac{4p^2}{q^2}\right)^{i\eta}e^{2i\eta\gamma}\frac{\Gamma(1+i\eta)}{\Gamma(1-i\eta)} \\
\label{f_c_int3}
& =& -\frac{2\eta}{q^2}\left(\frac{4p^2}{q^2}\right)^{i\eta} = -\frac{2\eta}{q^2}\left(\frac{1-\cos{\theta}}{2}\right)^{-i\eta},
\ea
where we have expanded the ratio of  gamma functions $\Gamma(1\pm i\eta)$ to first order in $\eta$. The result is analytic in $\eta$, and can be continued back to $\Im\,\eta=0$, giving the usual Coulomb amplitude, but with the relativistic rather than nonrelativistic value of $\eta$.

We can obtain the same result by introducing a convergence factor $e^{-ab}$ in the integrand for $\eta$ real and then using the second form of the result in \cite[Eq.\,13.2(3)]{Watson}. This gives
\be
\label{limit_form1}
\int_0^\infty db \,b^{1+2i\eta}e^{-ab}J_0(qb) = \frac{1}{(a^2+q^2)^{1+i\eta}}\Gamma(2+2i\eta)\, _2F_1\left(1+i\eta,-\frac{1}{2}-i\eta;1;\frac{q^2}{q^2+a^2}\right). 
\ee
Taking the limit $a\rightarrow 0$ using the limiting form of the hypergeometric function for unit argument and the duplication formula for gamma functions reproduces the result in \eq{f_c_int2}.

Cahn \cite{Cahn} obtained a similar result with the factor $(2p/q)^{2i\eta}$ replaced by $(\lambda/q)^{2i\eta}$  through somewhat loose arguments by starting with a screened Coulomb interaction with $1/q^2\rightarrow 1/(q^2+\lambda^2)$ in Born approximation. The difference is a phase factor $e^{2i\eta\ln{2p/\lambda}}$ with a phase which diverges for $\lambda\rightarrow 0$. This is a standard problem. The same phase appears in all terms in \eq{f^tot2}, so the phase does not affect the cross section and can be absorbed if it is done consistently in \eq{C+N_amp}. This phase is model-dependent in general. Bethe, for example, uses a Gaussian cutoff in \cite[Eq. 4.28 ff.]{BetheCoulomb}. Islam \cite{Islam} used a similar form to evaluate the Coulomb amplitude. The results in both cases differ from that in \eq{f_c_int2} only by (different) infinite phases in the Coulomb limit.

As noted above, magnetic-moment scattering does not contribute to the spin-independent part of the scattering amplitude. Its contribution to the spin-dependent amplitude is also suppressed  by an angular factor proportional to $\sqrt{q^2}$ at small $q^2$. The factor $1/q^2$ from the photon propagator therefore partially cancels and the amplitude is suppressed relative to the charge-scattering term for $q^2\rightarrow 0$.  The magnetic form factors $F_M^2(q^2)$  further suppress the scattering at large $q$. As a result, the magnetic terms  do not contribute significantly to the scattering in the region of interest here. This result is general, and holds also for scattering through higher multipoles in the case of particles of higher spin \cite[Secs.\, 4b,\,c]{VertexFuncs}. The potential interference effects of the magnetic terms with spin-dependent terms in the nuclear scattering studied in \cite{Buttimore} are also expected to be negligibly small at high energies as discussed in the Appendix.


\subsection{Form-factor corrections to Coulomb scattering \label{subsec:formfactors}}


We turn now to the evaluation of the second integrals in \eq{C+N_amp} which contains the effects of the form factors on the Coulomb scattering. We will use the standard form
\be
\label{form_factors}
F_Q(q^2) =\frac{\mu^4}{(q^2+\mu^2)^2}
\ee
for the proton charge form factor with $\mu^2=0.71$ GeV$^2$. We will write the $q^2$-dependent factors in the initial Born amplitude as
\ba
\frac{\alpha}{q^2}F^2(q^2) &=& \frac{\alpha}{q^2} - \frac{\alpha}{q^2}\left(1-\frac{\mu^8}{(q^2+\mu^2)^4}\right) \nonumber \\ 
\label{form_factors2}
&=& \frac{\alpha}{q^2}-\sum_{m=0}^3\alpha\frac{\mu^{2m}}{(q^2+\mu^2)^{m+1}}.
\ea
Applying an inverse Bessel transform to the second term, which we identify as the leading term in an expansion of the eikonal in terms of $2\delta_c^{FF}$, gives 
\be
\label{FF_eikonal}
2\delta_c^{FF}=-\sum_{m=1}^4  \frac{\alpha}{2^m\Gamma(m+1)}(\mu b)^mK_m(\mu b)
\ee
as the eikonal function for the form-factor corrections, where we have used the result in  \cite[Eq.\,10.22.46]{DLMF} and the symmetry $K_{-\nu}(z)=K_\nu(z)$ in evaluating the integrals. 

Substituting the form-factor term in \eq{C+N_amp} and expanding to leading order in $\delta_c^{FF}$, allowed because the form-factors terms are small and compact in impact-parameter space, we obtain a sum of integrals of the form
\ba
&& \frac{1}{2^m\Gamma(m+1)} e^{2i\eta\gamma}\int_0^\infty
db\,b (pb)^{2i\eta}(\mu b)^mK_m(\mu b)J_0(qb)   \nonumber \\
\label{FFint}
&&\quad  =\frac{1}{\mu^2}\left(\frac{2p}{\mu}\right)^{2i\eta}e^{2i\eta\gamma}\frac{\Gamma(m+1+i\eta)\Gamma(1+i\eta)}{\Gamma(m+1)}\, _2F_1\left(m+1+i\eta,1+i\eta;1;-\frac{q^2}{\mu^2}\right) \\
\label{FFint2}
&&\quad  =\left(\frac{4p^2}{q^2+\mu^2}\right)^{i\eta}e^{2i\eta\gamma} \frac{\mu^{2m}}{(q^2+\mu^2)^{m+1}}\frac{\Gamma(m+1+i\eta)\Gamma(1+i\eta)}{\Gamma(m+1)}\, _2F_1\left(m+1+i\eta,-i\eta;1;\frac{q^2}{q^2+\mu^2}\right),
\ea
where the result follows from \cite[Eq.\,13.45(1)]{Watson} and a standard linear transformation on the resulting hypergeometric function. 

The parameter $\eta$ is very small. Setting it equal to zero in the remaining hypergeometric function, expanding appropriately elsewhere, and summing over $m$ we find that the form-factor correction to the leading Coulomb amplitude is
\be
\label{FFint3}
-\sum_{m=0}^3 \alpha \left(\frac{4p^2}{q^2+\mu^2}\right)^{i\eta}\frac{\mu^{2m}}{(q^2+\mu^2)^{m+1}} = -\frac{\alpha}{q^2} \left(\frac{4p^2}{q^2+\mu^2}\right)^{i\eta}\left(1-\frac{\mu^8}{(q^2+\mu^2)^4}\right)\left[1+O(\eta)\right].
 \ee
Supplying the the necessary overall factors and adding the unmodified Coulomb term, the full Coulomb amplitude with the form-factor corrections  becomes
 \be
 \label{CoulFF}
 f_c(s,q^2)+f_c^{FF}(s,q^2) = -\frac{2\eta}{q^2}\left(\frac{4p^2}{q^2}\right)^{i\eta}\left[1 -\left(\frac{q^2}{q^2+\mu^2}\right)^{i\eta}  + \left(\frac{q^2}{q^2+\mu^2}\right)^{i\eta}\frac{\mu^8}{(q^2+\mu^2)^4}\right].
 \ee

 The form-factor corrections do not have the simple form of the pure Coulomb amplitude multiplied by $F^2(q^2)$ assumed, for example, by Cahn \cite{Cahn} when the phase factors are included. However, the limiting behaviors of the full amplitude are evident from \eq{CoulFF}. For $q^2\ll\mu^2$, the last two terms in this expression cancel, and the full result approaches the pure Coulomb amplitude as expected from the infinite range of that interaction. In the opposite limit, $q^2\gg\mu^2$, the first two terms eventually to cancel to an extra $O(\eta)$, while the last term approaches the pure Coulomb result multiplied by $F^2(q^2)$.  All terms are important for intermediate $q^2$.
 
 We can see the structure of \eq{CoulFF} more clearly by looking separately at the amplitude and phase of the factor which multiplies the Coulomb amplitude $-\frac{2\eta}{q^2}e^{i\eta \ln{4p^2/q^2}}$:
 \ba
 \label{fcFFmag}
 {\rm mag}[\cdot] &=& \frac{ \mu^8}{(q^2+\mu^2)^4}\left[1+O(\eta^2)\right]  \\
 \label{fcFFarg}
{\rm arg}[\cdot] &=& - \arctan{\left[\eta\left(\frac{(q^2+\mu^2)^4}{\mu^8}-1\right)\ln{\frac{q^2}{q^2+\mu^2}}\right]} \\
&=&- \eta\left(\frac{(q^2+\mu^2)^4}{\mu^8}-1\right)\ln{\frac{q^2}{q^2+\mu^2}}\left[1+O\left(\eta^2\frac{(q^2+\mu^2)^4)}{\mu^8}\right)\right].
 \ea
 The approximation in the last line is valid in the region of small $q^2$ where Coulomb-nuclear interference is significant, with $q^2<\mu^2$.
 
 We remark that other parametrizations of the proton charge form factor consistent with the dispersion relations for that quantity --- expressions in involving sums or integrals of  inverse powers of quantities  $(q^2+\lambda^2)$ --- lead to results of the same general form, but involving further sums or integrals.  The differences among the common parametrizations are unimportant. 
 
 Using the results in Eqs.\ (\ref{CoulFF}), (\ref{fcFFmag}), and  (\ref{fcFFarg}), we obtain finally for the form-factor corrections to the Coulomb amplitude for $\eta/F_Q^2(q^2)\ll 1$
 \ba
 \label{CoulFF2}
 f_c(s,Q^2)+f_c^{FF}(s,q^2) &=& -\frac{2\eta}{q^2}F_Q^2(q^2)e^{i \Phi_{c,FF}}, \\
 \label{PhicFF}
\Phi_{c,FF}(s,q^2) &\approx& \eta\ln{\frac{4p^2}{q^2}}-\eta \left(\frac{(q^2+\mu^2)^4}{\mu^8}-1\right)\ln{\frac{q^2}{q^2+\mu^2}},
 \ea
 where we have used the (standard) parametrization for charge form factor in \eq{form_factors}.

 
 \subsection{Coulomb and form-factor corrections to the nuclear amplitude \label{subsec:CoulNuclear}}
 
 
We turn next to the final term in \eq{f^tot2}, the amplitude for the nuclear scattering including the effects of the Coulomb phase shifts and the form-factor corrections,
\be
\label{fNc}
f_{N,c}(s,q^2) =  +i\int_0^\infty db\,b\,e^{2i\delta_c(b,s)+2i\delta^{FF}_c(b,s)}\left(1-e^{2i\delta_N(b,s)}\right)J_0(qb),
\ee
where $\delta_c(b,s)=\eta\ln{pb}+\eta\gamma$ and $\delta^{FF}_c(b,s)$ is given in \eq{FF_eikonal}. The form-factor effects can be isolated subject to Coulomb and nuclear corrections by expanding to first order in $\delta_c^{FF}$, but we have not found this to be especially useful. The Coulomb effects are of course long range, and a similar expansion of those terms is not useful. 

A simple approximation to the integral in \eq{fNc} is to replace the Coulomb and form-factor phases in that expression by their values at the peak of the uncorrected eikonal distribution, and factor the resulting constant phase out of the integral. This approximation, originally suggested by Bethe \cite{BetheCoulomb}, works well for the slowly-varying Coulomb factor. It is less accurate when the form-factor term is included as that expression varies significantly over the same range in impact parameter as the eikonal term itself, and affects the value of the integral. The dependence of the Bessel function $J_0(q b)$ in \eq{fNc} also introduces strong dependence on $q^2$ as that quantity increases, leading to diffraction zeros in the real and then the imaginary parts of the amplitude, so the Bethe approximation is only valid at small $q^2$. 

We note for completeness that the location of the peaks in both the imaginary and real parts of the nuclear distribution is at  $b_{\rm peak}\approx\sqrt{\sigma_{\rm tot}/4\pi}$ at $q^2=0$ \cite{eikonal2015}. The $q^2$ dependence of the Bessel function, $J_0(qb)= 1-\frac{1}{4}q^2b^2+\cdots$ introduces a term proportional to $b^2$  in first order, and shifts the peak to a location determined by the logarithmic slope parameter $B$, with
\be
\label{F_NCapprox}
f_{N,c}(s,q^2) \approx \left(e^{2i\delta_c(b,s)+2i\delta_c^{FF}(b,s)}\Big|_{b=\sqrt{\sigma_{\rm tot}(s)/4\pi}}+\frac{1}{2}q^2B\,e^{2i\delta_c(b,s)+2i\delta_c^{FF}(b,s)}\Big|_{b=\sqrt{B(s)}}\right)f_N(s,q^2) +O(q^4)
\ee
in the Bethe approximation.

Rather than using this approximation, our approach has instead been to use our existing eikonal fit to $pp$ scattering \cite{eikonal2015}, evaluate the integral in \eq{fNc} directly, and then compare the results to a direct evaluation of the nuclear amplitude itself, \eq{f_N}. Any fit to the $pp$ scattering amplitude consistent with unitarity can be put in eikonal form, and a similar analysis made. However, we stress that successful fits cannot deviate substantially from our fit, which describes the scattering reasonably well from 5 GeV to the TeV range.  In particular, we calculate the modified amplitude in \eq{fNc}  numerically using the eikonal model in \cite{curvature_fit} and relate the result in phase-amplitude form to the pure nuclear amplitude.

\begin{figure}[htbp]
\includegraphics{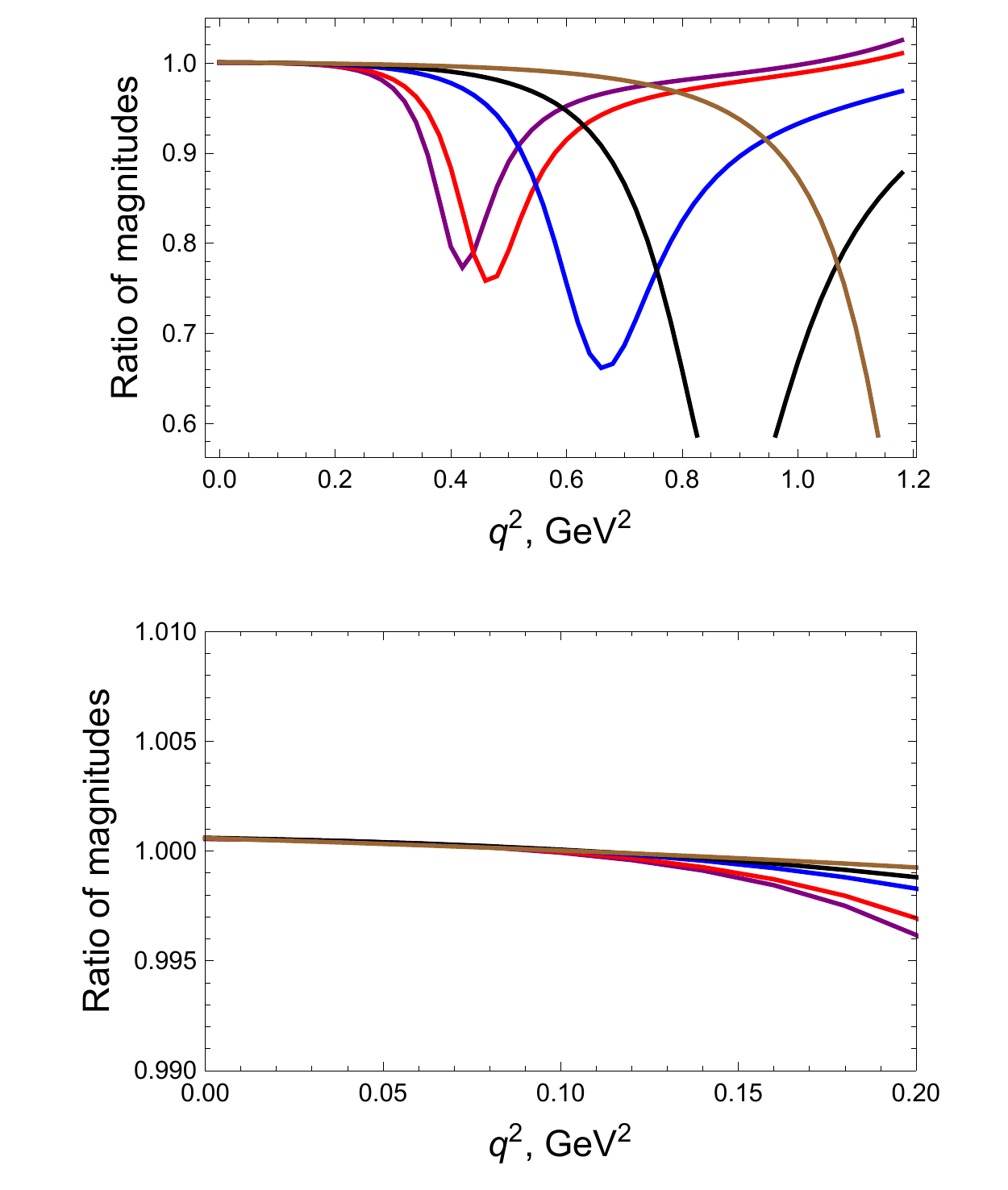}
 \caption{The ratios of magnitudes $\lvert f_{Nc}(s,q^2)\rvert\big/\lvert f_N(s,q^2)\rvert$ at 100 (brown), 546 (black), 1800 (blue), 8000,  (red) and 13000 (purple) GeV, top to bottom on the right-hand side of the lower figure, top to bottom on the left in the upper figure.}
 \label{fig:ratios}
\end{figure}

Our results for the ratio $\lvert f_{N,c}(s,q^2)/f_N(s,q^2)\rvert$  of the ratio of  magnitudes of the nuclear amplitude with Coulomb and form-factor corrections to the pure nuclear amplitude is shown in \fig{fig:ratios}. As expected, and seen in the upper half of the figure, the ratio is very close to unity at small $q^2$ where the amplitudes should differ by a pure phase in the Bethe approximation, but differs substantially at large $q^2$ where that approximation fails and the diffraction structure of the amplitudes is important. As seen in the lower half of the figure, the corrections are a small fraction of a percent for $q^2<0.2$ GeV$^2$, the  range  used in typical analyses, and less that 0.1\% near $q^2=0$. We will ignore these corrections, and treat the ratio of amplitudes as a pure phase.

\begin{figure}[htbp]
\includegraphics{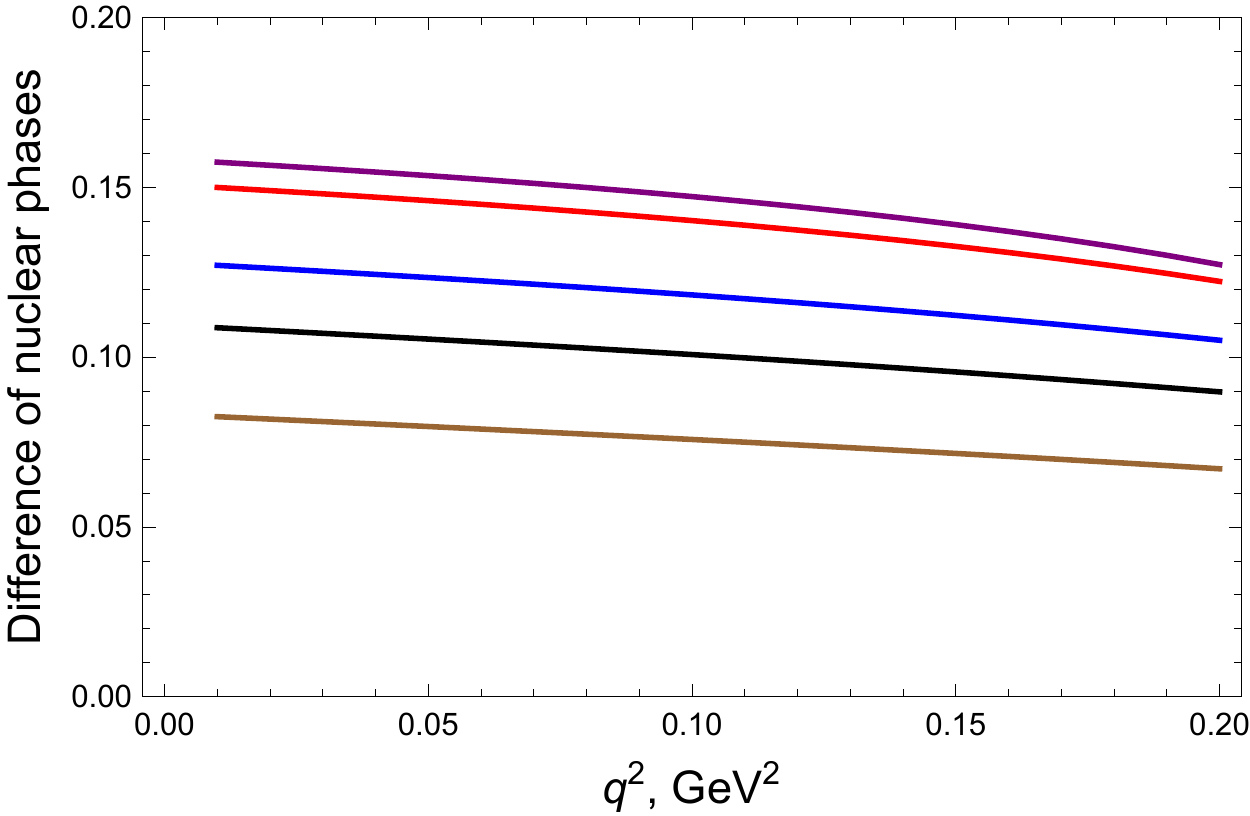}
 \caption{The differences of nuclear phases $\Delta\Phi_N(s, q^2)=\arg\left(f_{N,c}(s,q^2)\right)-\arg\left( f_N(s,q^2)\right)$ at (top to bottom) at 13000 (purple), 8000 (red), 1800 (blue), 546 (black), and 100 (brown)  GeV.}
 \label{fig:nucl_phases}
\end{figure}

We show the difference $\Delta\Phi_N(s, q^2)=\Phi_{N,c}-\Phi_N$ between the phases of the corrected and pure nuclear amplitudes, 
\be
\label{phase_diff}
\Delta\Phi_N(s,q^2) = {\rm arg}f_{N,c}(s,q^2)-{\rm arg}f_N(s,q^2)={\rm arg}\left(f_{N,c}(s,q^2)/f_N(s,q^2)\right)
\ee
 in \fig{fig:nucl_phases}. The energy  dependence of this difference is due mostly to the factor $(pb)^{2i\eta}$ in the Coulomb phase in \eq{fNc}; the same dependence on $p$ appears in the Coulomb and form-factor term in \eq{CoulFF}, and will drop out in the differential cross section. 
  
 The dependence of $\Delta\Phi_N(s, q^2)$ on $s$ and $q^2$ is described very well by an expression quadratic in $q^2$ and linear in the momentum $p(W)$,
 \be
 \label{phase_fit}
\Delta \Phi_N(s,q^2)\approx a_1+a_2 q^2+a_3q^4 + (b_1+b_2q^2+b_3q^4)\log{p}.
 \ee
 The parameters in the fit are given in Table \ref{table1:phasefit}.


\begin{table}[ht]                   
%
\def\arraystretch{1.15}            
\begin{center}				  
\begin{tabular}[b]{|c|c|}

\hline
{\rm Parameter} & Value, radians \\
\hline
       $a_1$\ \   &\  0.02346\ \\
       $a_2$ \ \  &\ -0.08661\ \\
       $a_3$ \ \ &\  0.34517\   \\
       $b_1$ \ \ &\  0.01530\ \\
       $b_2$ \ \ &\ 0.00280\ \\
       $b_3$ \ \ &\ -0.08678\ \\
\hline
	
\end{tabular}
     \caption{The parameters in the fit in \eq{phase_fit} to the phase difference   \label{table1:phasefit}}
\end{center}
\end{table}
\def\arraystretch{1}  

\begin{figure}[htbp]
\includegraphics{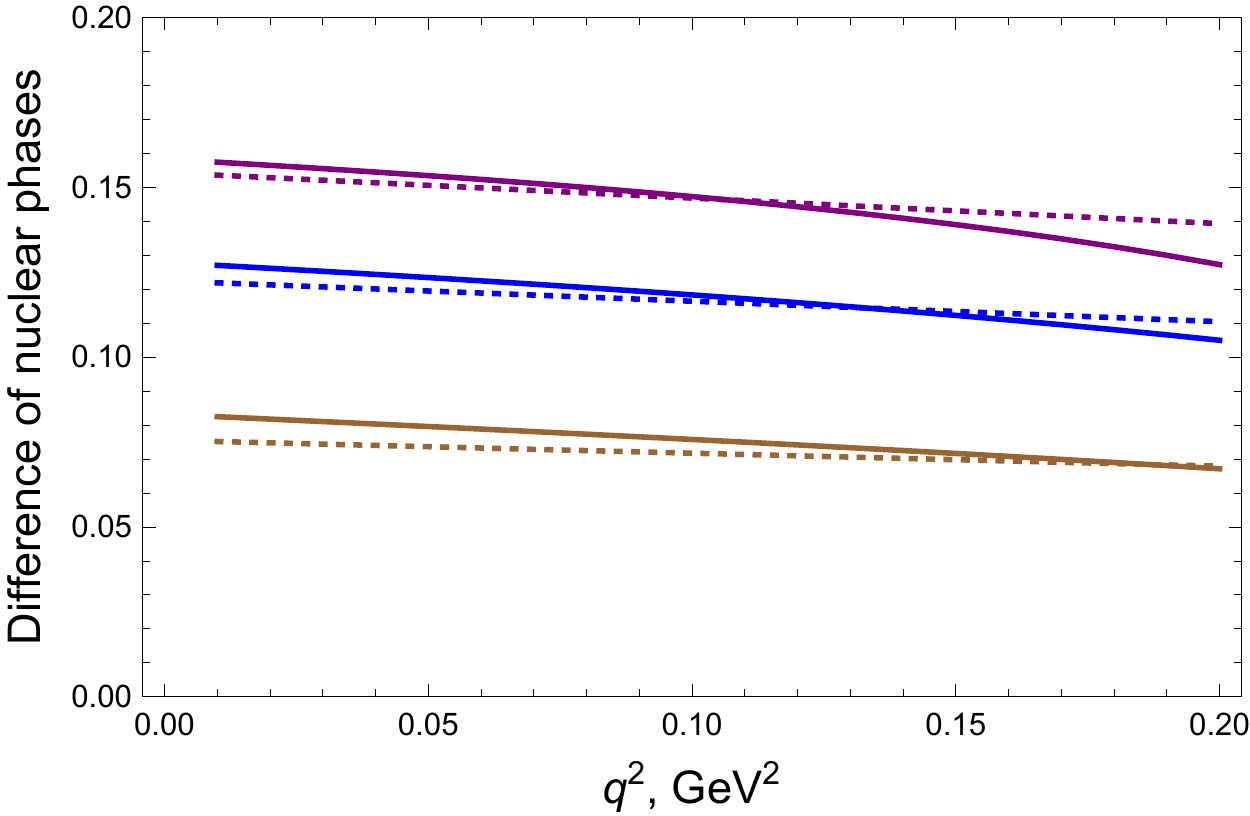}
 \caption\protect{Phases in the Bethe approximation of \eq{F_NCapprox} (dashed curves) compared to the exact phases (solid curves)  at, top to bottom, 13,000 (purple), 1800 (blue), and 100 (brown) GeV.}
 \label{fig:Bethe}
\end{figure}

\begin{figure}[htbp]
\includegraphics{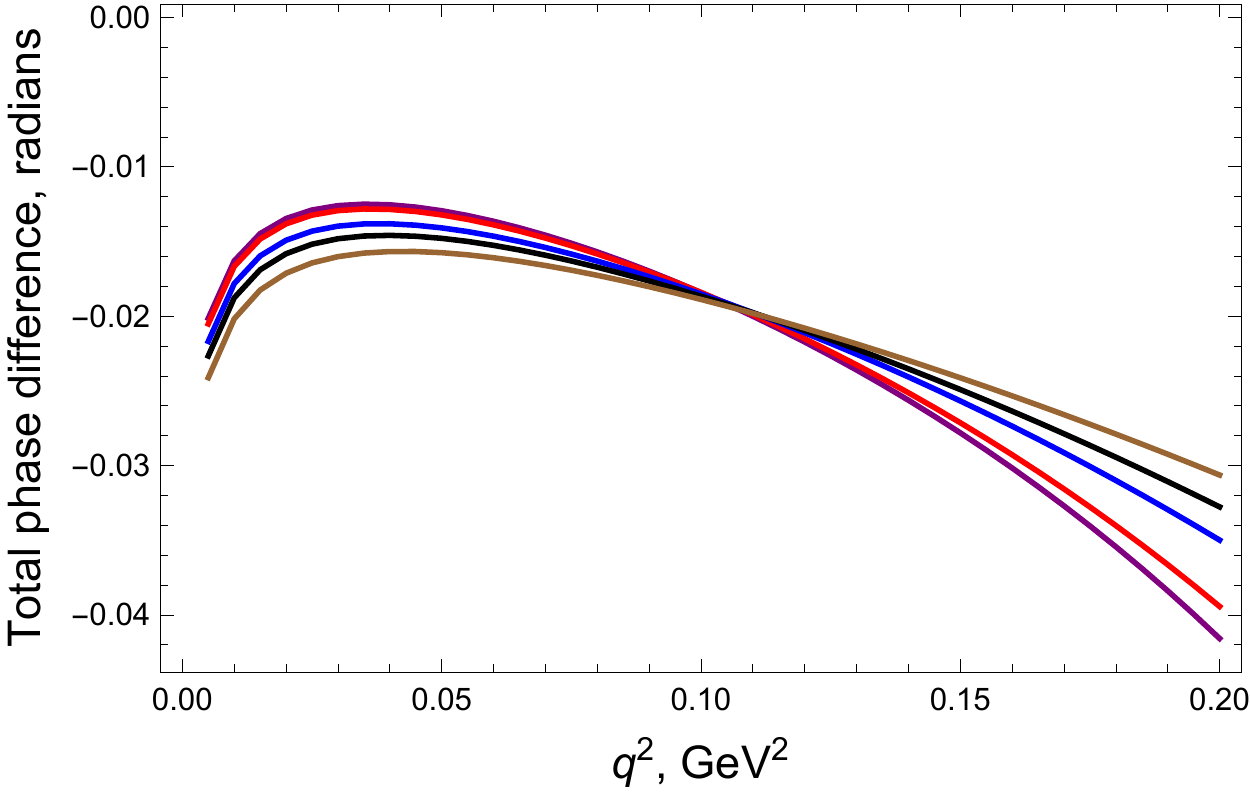}
 \caption\protect{Total phase differences $\Phi_{tot}(s,q^2)$, \eq{totArgdiff}, between the nuclear and Coulomb amplitudes at 100 (brown), 546 (black), 1800 (blue), 8000 (red), and 13000 (purple) GeV, top to bottom on the right.}
 \label{fig:total_phases}
\end{figure}

 We compare these results with those obtained in the Bethe approximation in \fig{fig:Bethe}. While differences do not appear to be large, they are significant on the scale of  the final phase differences in \fig{fig:total_phases}. In that figure, we show the total phase difference $\Phi_{tot}(s,q^2)$ between the Coulomb and nuclear parts of the scattering amplitude when it is written so that the Coulomb term is real with the phase $(4p^2/q^2)^{i\eta}$ absorbed,
 \ba
 \label{final_amp2}
 f(s,q^2)&=& -\frac{2\eta}{q^2}F^2(q^2) + e^{i\Phi_{tot}(s,q^2)}f_N(s,q^2), \\
 \label{totArgdiff}
 \Phi_{ tot}(s,q^2) &=& \Phi_{N,c}(s,q^2)-\Phi_N(s,q^2)-\Phi_{c,FF}(s,q^2).
 \ea
The main energy dependence of $\Phi_{N,c}(s,q^2)$ through the factor $p^{2i\eta}$ in the integrand for $f_{N,c}(s,q^2)$, \eq{fNc}, cancels with the corresponding factor in $\Phi_{c,FF}$; the residual energy dependence arises from that in the nuclear part of the integrands.

The form of the amplitude in \eq{final_amp2} is particularly convenient. The Coulomb amplitude is real, so the only source of Coulomb-nuclear interference in the differential cross section is through the real part of the nuclear term. The phase $\Phi_{tot}$ is quite small as seen in \fig{fig:total_phases}, so the main effect of the phase factor $e^{i\Phi_{tot}(s,q^2)}=1+i\Phi_{tot}(s,q^2)+\cdots$ is to mix a small component of $\Im f_N(s,q^2)$ into the real part $\Re f_N(s,q^2)$. While fractionally small, it is still significant because $\Im f\gg\Re f$ in the interference region. This structure is not immediately evident when the Coulomb and nuclear phases are included separately on those terms. The change in the imaginary part of the amplitude is small, given by the product of $\Re f_N$ and $\Phi_{tot}$.

Separating out these correction terms, we can write the complete amplitude as
\be
\label{fcorrected}
 f(s,q^2) = -\frac{2\eta}{q^2}F^2(q^2) - \Phi_{tot}(s,q^2)\Im f_N(s,q^2)+i\Phi_{tot}(s,q^2)\Re f_N(s,q^2)   +f_N(s,q^2).
 \ee
 In this form, the pure nuclear amplitude is  displayed separately. However, it is important to recognize that the corrections will affect any attempt to determine $f_N$ by fitting data away from the main Coulomb-nuclear interference region, and must be taken into account, as must the tail of the Coulomb term.

\begin{figure}[htbp]
\includegraphics{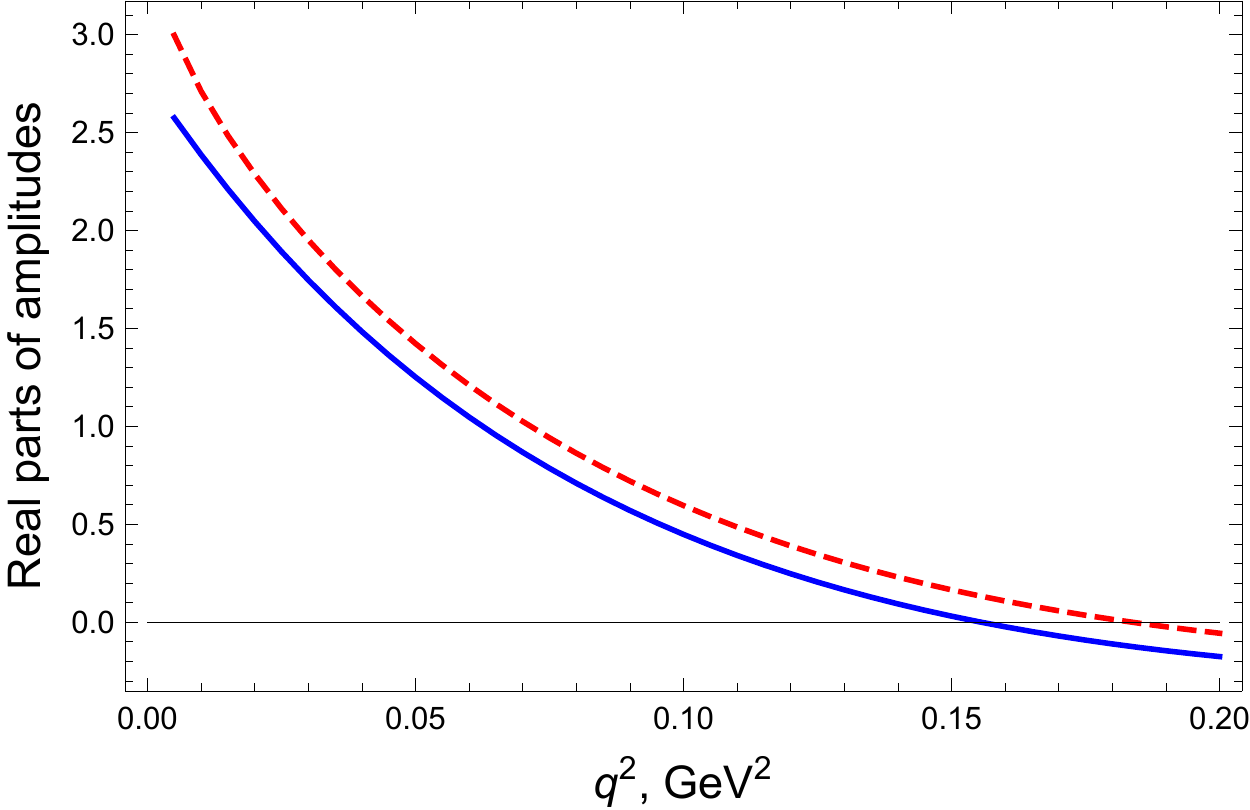}
 \caption\protect{Comparison of the real part of the actual nuclear amplitude at 13 TeV (solid blue curve) with the effective real part including the Coulomb and electromagnetic form-factor corrections, Eqs.\  (\ref{final_amp2}) and (\ref{totArgdiff}), (dashed red curve).}
 \label{fig:effRefNcomp}
\end{figure}

The effect of the correction is illustrated in \fig{fig:effRefNcomp}, where we compare the real parts of the nuclear amplitude with and without the phase correction. The final correction is not large. A reasonable first approximation is in fact to ignore the correction and take the full amplitude simply as the sum of the (real) leading-order Coulomb term with form factors and the uncorrected nuclear amplitude. However, in the present form, the corrections are very simple to include, and should be used. The simplicity is striking relative to the treatment of the corrections in \cite{Cahn,KL-Coulomb}.

We emphasize that the corrections shown in \fig{fig:total_phases} are very stable and do not change for reasonable changes in the eikonal model. This is to be expected: in the Bethe approximation, the corrections to the nuclear phase are independent of the details of the model, and require only that the eikonal amplitude be strongly peaked in impact parameter space. This is a generic feature of realistic scattering amplitudes at high energies. We note also that the corrections depend mostly on the imaginary part of the nuclear impact parameter distribution, which is dominant at high energies and well determined by fits to the total cross section and the slope parameter $B$. We give an example of an alternative model in the next section; the changes in the corrections are indiscernible in the equivalent of \fig{fig:total_phases}.

\begin{figure}[htbp]
\includegraphics{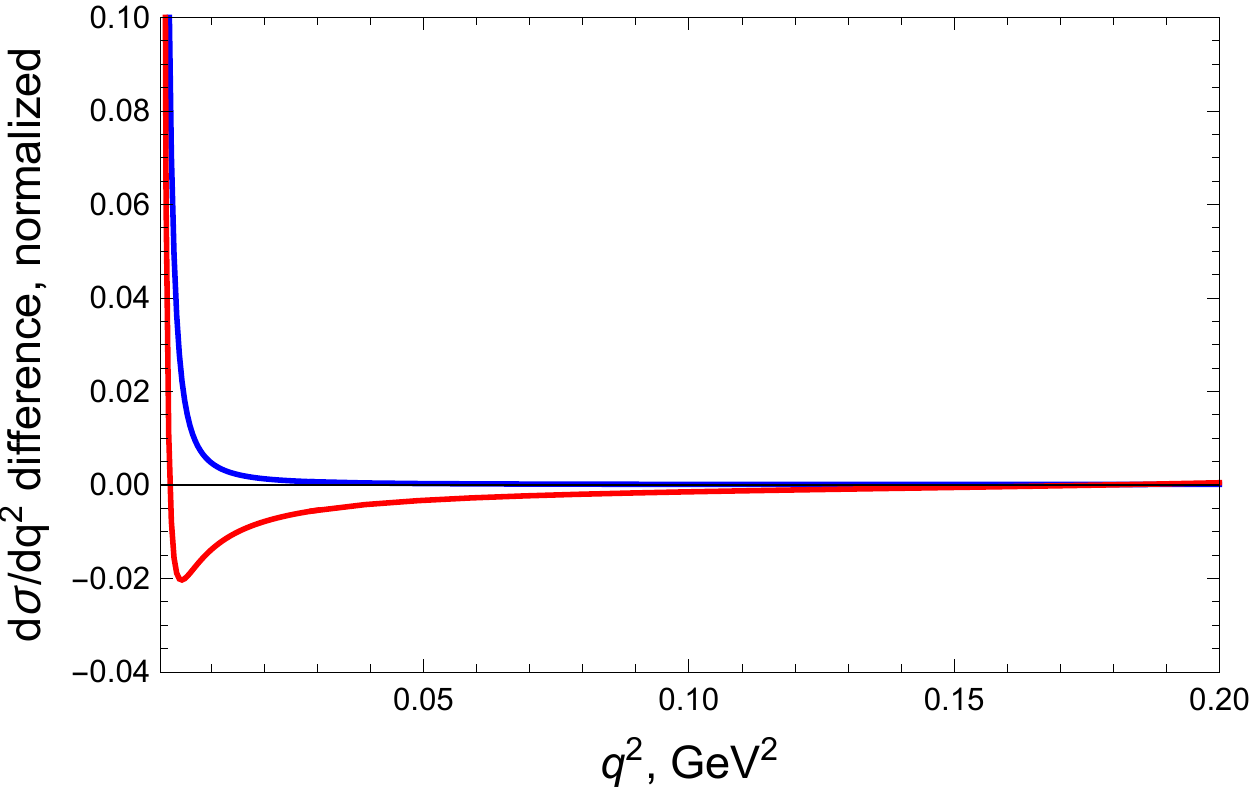}
 \caption{Plots of the cross section ratio $\left(d\sigma/dq^2-d\sigma_N/dq^2\right)/\left(d\sigma_N/dq^2\right)$ (lower red curve) and the corresponding ratio with the Coulomb-nuclear interference term dropped (upper blue curve) in $pp$ scattering in the eikonal model at 13 TeV. }
 \label{fig:dsigmacomp}
\end{figure}

In \fig{fig:dsigmacomp} we show the effects of the Coulomb-nuclear interference on the final cross section, plotting the ratio
$\left(d\sigma/dq^2-d\sigma_N/dq^2\right)/\left(d\sigma_N/dq^2\right)$ as a function of $q^2$ at 13 TeV (lower red curve). We also show the ratio of cross sections in the absence of interference, $\left(d\sigma_c/dq^2+d\sigma_N/dq^2\right)\big/\left(d\sigma_N/dq^2\right)-1=\left(d\sigma_c/dq^2\right)\big/\left(d\sigma_N/dq^2\right)$ (upper blue curve), where $d\sigma_c/dq^2$ is the Coulomb cross section. The interference effects are small, at most $\sim 2\%$ at the dip at $q^2=0.0045$ GeV$^2$, that is, of order $\eta\approx\alpha$, and are significant only at very small values of $q^2$. This is the result of the $1/q^2$ fall-off of the Coulomb amplitude further suppressed by the effects of the proton charge form factors. For comparison, the statistical uncertainties in the measured cross sections in \cite{TOTEM2016} vary from $\sim 0.5\%$ at the dip to $\sim 1\%$ for $q^2\approx 0.2$. The results are similar at lower energies, with the only significant sensitivity in the region of the dip.

\section{An application and simple model for the amplitudes \label{sec:application}}


As an application of these results, we consider a model which has been used frequently in the analysis of experimental data, {\em e.g.}, the TOTEM data at 8 and 13 TeV; see \cite{TOTEM2016,TOTEM2019} and earlier references therein. In this model, the phase of the nuclear amplitude is taken as a constant independent of $q^2$. It is determined simply by the ratio  $\rho$ of the  real to the imaginary parts of the nuclear amplitude in the forward direction, corresponding to a phase $\Phi_N(s,q^2)\equiv\frac{\pi}{2}-\arctan{\rho(s)}$ and $f_N(s,q^2)=e^{i\Phi_N}\lvert f_N(s,q^2)\rvert$. This is clearly unrealistic in general, but may be adequate in a small region near $q^2=0$. 

We will consider two versions of this model. In the first, we write  the complex eikonal amplitude in \eq{f^tot} approximately in terms of its dominant imaginary part multiplied by a constant phase and properly normalized,
\be
\label{const_phase_eikonal}
i\left(1-e^{2i(\delta_c^{tot}+\delta_N)}\right) \longrightarrow ie^{-i\arctan{\rho}}\,\Im\,i\left(1-e^{2i\left(\delta_c^{tot}+\delta_N\right)}\right) /\cos\left(\arctan{\rho}\right).
\ee
This form allows us to calculate the Coulomb and form-factor corrections to the nuclear phase. As discussed above, the results are essentially identical to those obtained using the full amplitude shown in \fig{fig:total_phases} even though  the real  parts of the amplitudes and the cross sections differ. This is as expected given the Bethe argument. 

Since the corrections are effectively model-independent, we can proceed to a simpler construction used in various experimental analyses and write the constant-phase amplitude approximately in terms of the standard small-$q^2$ expansion of the nuclear scattering cross section,
\be
\label{curvature_exp}
\frac{d\sigma}{dq^2}(s,q^2) \approx Ae^{-Bq^2+Cq^4-Dq^6+\cdots},\quad 0\leq q^2\ll 1,
\ee
where $B$ is the usual slope parameter and $C,\,D\,\cdots$ introduce curvature in $d\sigma/dq^2$. Taking the square root and introducing a phase, we have
\be
\label{fNcurv}
\sqrt{\pi}f_N(s,q^2) \approx \sqrt{A}\,e^{i\Phi_N}e^{-\frac{1}{2}\left(Bq^2-Cq^4+Dq^6-\cdots\right)}.
\ee

We will initially take $\Phi_N$ as constant, with $\Phi_N=\frac{\pi}{2}-\arctan{\rho}$. This is the form assumed, for example, in the TOTEM analyses of Coulomb-nuclear interference \cite{TOTEM2016,TOTEM2019}, with $\rho$ used as a parameter in fitting the data in the interference region. Note that this form, with $\Phi_N$ constant, does not allow for zeros and the associated changes in sign of the real and imaginary parts of the amplitude as at the diffraction zeros in $f_N$, so is restricted to small $q^2$. 

The expansion in \eq{curvature_exp} and its range of validity were investigated in detail in \cite{bdhh-curvature}, where exact expressions were given for the parameters $B,\,C,\,{\rm and}\ D$ in the eikonal approach. As noted there, the predicted values of those parameters were consistent with the results obtained by TOTEM Collaboration in their fits to their TeV data \cite{TOTEM2015}.  The next term in the series becomes important near the upper end of the range of $q^2$ used in the TOTEM fits, with errors in the fitted cross section comparable to, or larger than, the uncertainties in the experimental results. 

In general, fits based on \eq{curvature_exp} should use $q^2\lesssim 0.1-0.15$ GeV$^2$ at the higher energies; TOTEM used values of $q^2$ up to 0.2 GeV$^2$. This use of a too-wide range of $q^2$ is also common in analyses at lower energies; corrections to the quoted results were considered in detail in \cite{curvature_fit}.

It is straightforward to estimate the value of the next coefficient in the series using the calculated value of the local slope parameter $B(q^2)$ \cite{bdhh-curvature} at a small value of $q^2$ such as $q_0^2=0.01$ GeV$^2$. Since $B(q_0^2)=-d \log({d\sigma/dq^2})/dq^2\rvert_{q_0^2}\approx B-2Cq_0^2+3Dq_0^4-4Eq_0^6+O(q_0^8)$, we can express $E$ in terms of $B(q_0^2)$ and the known values of $B,\,C,\,{\rm and}\  D$. We will not use this refinement here, though it extends the range of validity of the expansion to approximately that used in \cite{TOTEM2015,TOTEM2016,TOTEM2019}, but will follow the procedures used there and simply fit the exact eikonal amplitudes at 8 and 13 TeV using the expression in \eq{fNcurv}. The fitted values of $A$ and $B$ do not differ significantly from the exact values. $C$ changes by a few percent, and $D$ changes significantly. The results are consistent with those found by the TOTEM Collaboration.

\begin{figure}[htbp]
\includegraphics{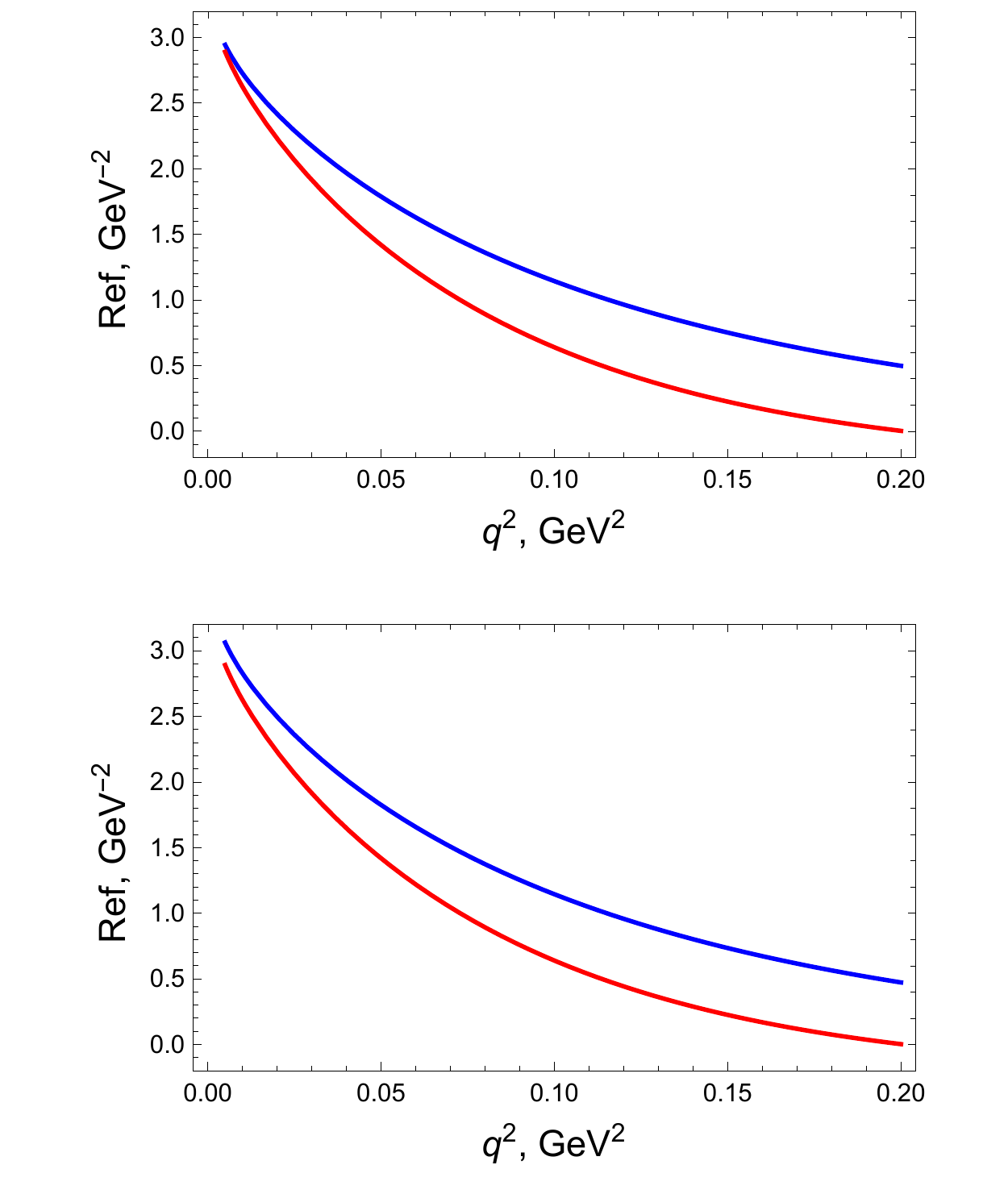}
 \caption{Comparison of the real part of the nuclear amplitude with Coulomb and form-factor corrections included (bottom red curves) with the corresponding real part of the amplitude in the constant phase approximation (top blue curves)  at W = 8 TeV (top figure) and  13 TeV (bottom figure).}
 \label{fig:Refcomp}
\end{figure}

We compare the results for $\Re f_N(s,q^2)$ obtained using the fits and \eq{fNcurv} with the exact eikonal results at 8 and 13 TeV in \fig{fig:Refcomp}. The real parts  in the constant phase approximation (top blue curves) are systematically larger than the exact results (bottom red curves), suggesting that this approach will lead to reduced values of $\rho$ when used to fit data. A better approximation is needed.

As is evident from \fig{fig:effRefNcomp}, the real part of the nuclear amplitude drops rapidly with increasing $q^2$, and actually changes sign in the region used in the TOTEM analyses. We therefore propose a simple approximation for the phase which takes this behavior into account. Since $\Phi_N(s,q^2)=\frac{\pi}{2}-\arctan{\rho(s,q^2)}$, we concentrate on $\rho(s,q^2)=\Re f_N(s,q^2)\big/\Im f_N(s,q^2)$. At high energies, $\Re f_N$ has a zero at small $q^2$, $\Re f_N(s,q_R^2)=0$. Similarly, $\Im f_N(s,q_I^2)=0$ at the first diffraction dip in $d\sigma/dq^2$ at $q_I^2>q_R^2.$\protect\footnote{The existence of the diffraction zeros in the  $pp$ cross section was first predicted over fifty years ago---in the eikonal context---in \protect\cite{LD-Lipes} and \protect\cite{Chou-Yang}, with diffraction effects on the spin polarization also noted in \protect\cite{LD-Lipes}.} Taking these zeros into account, we write
\be
\label{rho_zero_approx}
\rho(s,q^2) \approx \rho(s)\,\frac{1-q^2/q_R^2}{1-q^2/q_I^2}.
\ee
This form reduces to $\rho=\rho(s)$ at $q^2=0$ and has the proper zeros built in, with $\Phi_N(s,q^2)=\pi/2$  at $q_R^2$ and $f_N(s,q_R^2)$ purely imaginary, and $\Phi_N=0$ at $q_I^2$ and $f_N$ real. The value of $\rho$ can again be used as a fitting parameter.

\begin{figure}[htbp]
\includegraphics{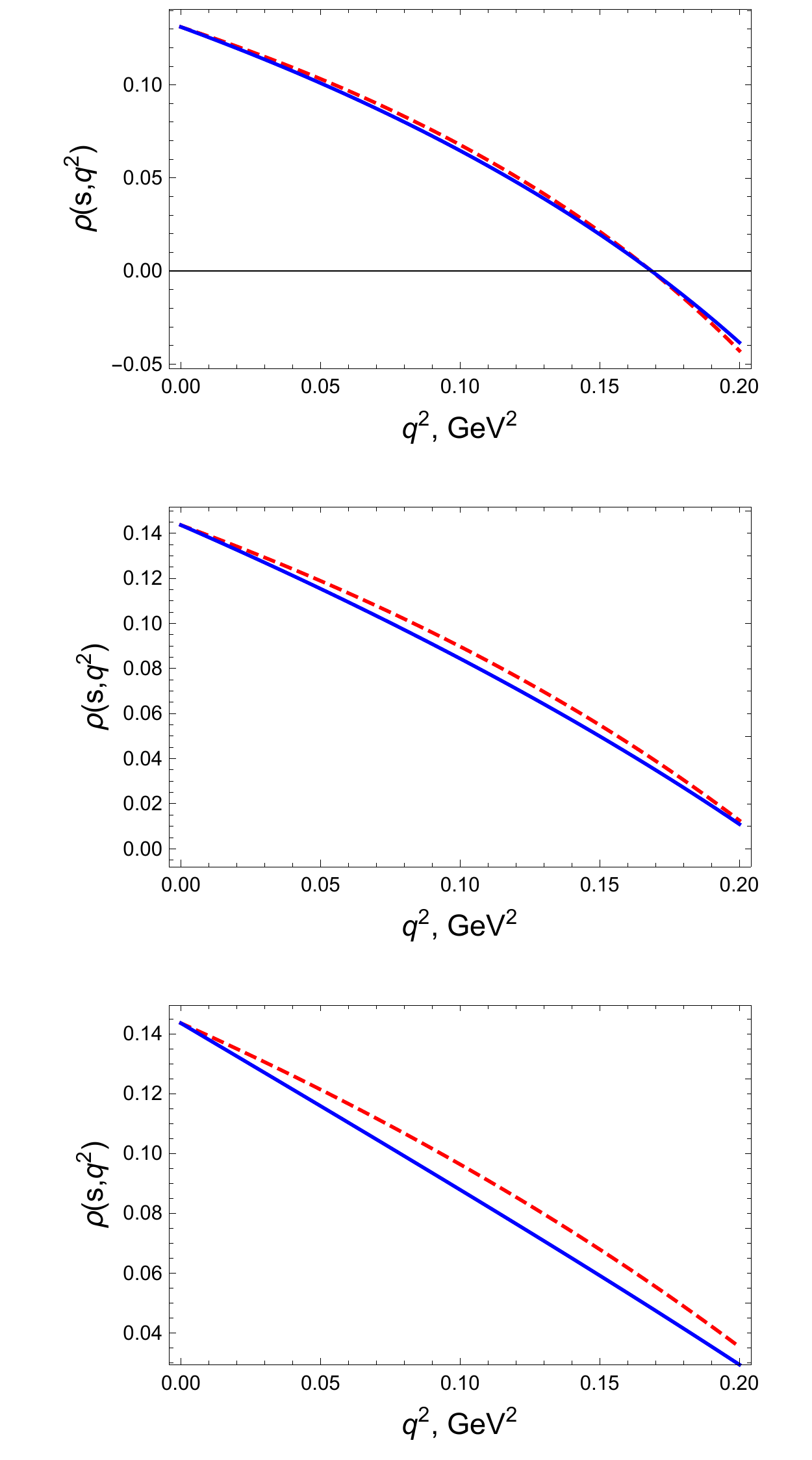}
 \caption{Comparison of the values of the ratio $\rho(q^2)=\Re f_N(s,q^2)/\Im f_N(s,q^2)$ calculated in the eikonal model at 8000, 1800, and 546 GeV, top to bottom, with the approximate values given in terms of the zeros of the real and imaginary parts of the amplitude by \eq{rho_zero_approx} . The actual values are given by the solid blue curves, the approximate values, by the dashed red curves.}
 \label{fig:RhoApprox}
\end{figure}

We compare the actual and approximate values of $\rho(s,q^2)$ in the eikonal model \cite{eikonal2015} at 8000, 1800, and 546 GeV in \fig{fig:RhoApprox}. The approximate results---and the corresponding results for $\Re f_N(s,q^2)$---are  quite accurate at the higher energies, and still good in the 500 GeV region. They are much better than the results obtained with the so-called "standard phase" used in some analyses which takes only the diffraction zero in the imaginary part of the amplitude into account \cite[Sec.\ 6.1.3]{TOTEM2016}, and then only approximately. 

We note that the errors in \fig{fig:RhoApprox} can be essentially eliminated by multiplying the expression in \eq{rho_zero_approx} by a factor $(1- aq^2)$ with an appropriate value of the coefficient $a$. This is useful in obtaining accurate fits to $\rho(s,q^2)$. However, the approximate expression in \eq{rho_zero_approx} requires knowledge only of the location of the zeros in $\Re f_N$ and $\Im f_N$. In particular, $q_I^2$ can be estimated from the diffraction structure of the cross section, while, roughly, $q_R^2\approx q_I^2/3$ at high energies in the eikonal model. 

\begin{figure}[htbp]
\includegraphics{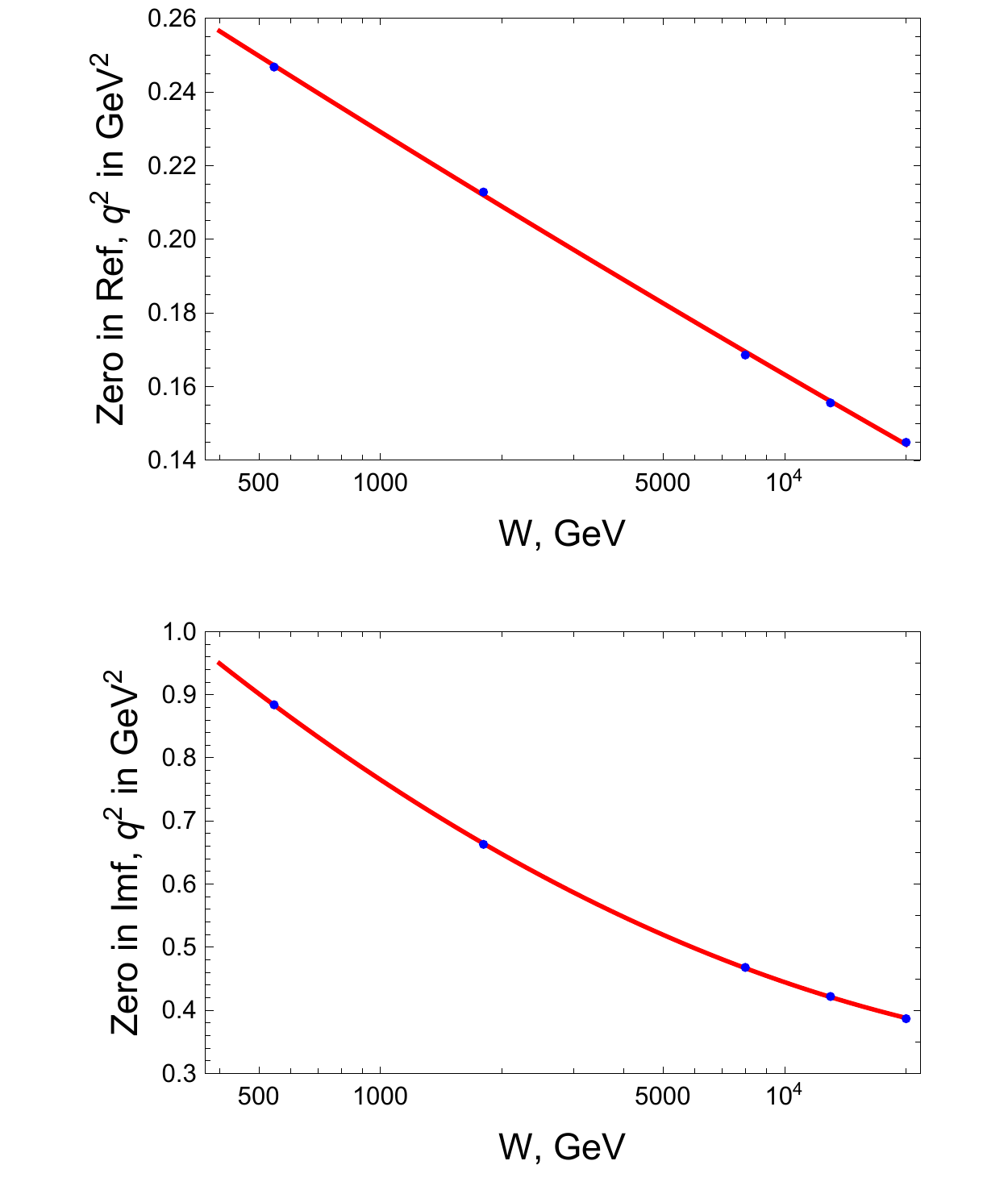}
 \caption{Location in $q^2$ of the zeros in the real and imaginary parts of the nuclear amplitude in the eikonal model as functions of $W$. The points give the actual values of $q^2$ at the zeros in the eikonal model; the  lines correspond to the fits in the text.}
 \label{fig:ZerosPlot}
\end{figure}

The actual location of the zeros in the real and imaginary parts of $f_N(s,q^2)$ in the eikonal model \cite{eikonal2015} are plotted in \fig{fig:ZerosPlot}. The curves in the figure correspond to a fit with
\ba
 q_R^2(W) &=& a_R+b_R \log{W}+c_R\log^2{W}, \nonumber \\
 \label{zeros_fit}
 q_I^2(W) &=& a_I+b_I \log{W}+c_I\log^2{W}.
 \ea
 The parameters in the fit are given in Table \ref{table2:zerofit}.  We note that the fit becomes inaccurate at energies below a few hundred GeV, where the zeros are displaced by small contributions from exchange terms dependent on inverse powers of $W$; see {\em e.g.}, \cite{eikonal2015}.

These results can be used in conjunction with the expansion in \eq{fNcurv} to construct nuclear amplitudes with a realistic $q^2$ dependence and phase at small $q^2$, again retaining $\rho=\rho(s)$ as a parameter to be used in fitting data. With the phase $\Phi_N=\frac{\pi}{2}-\arctan{\rho(s,q^2)}$ calculated using the expression for $\rho$ in \eq{rho_zero_approx}, the diffraction zeros in the real and imaginary parts of the amplitude are built in, and the magnitude of the amplitude is simply $\sqrt{d\sigma/dq^2}$.


\begin{table}[ht]                   
%
\def\arraystretch{1.15}            
\begin{center}				  
\begin{tabular}[b]{|c|c|}

\hline
{\rm Parameter} & Value, GeV$^2$ \\
\hline
      $a_R$\ \  & 0.4514   \\
      $b_R$ \ \  &-0.03484\\
       $c_R$ \ \ & 0.000386   \\
       $a_I$ \ \ & 2.9464 \\
       $b_I$ \ \ & -0.4481 \\
       $c_I$ \ \ & 0.01916  \\
\hline
	
\end{tabular}
     \caption{The parameters in the fit to the locations in $q^2$ of the zeros in the real and imaginary parts of the eikonal scattering amplitude in  $W$ given in \eq{zeros_fit}.  \label{table2:zerofit}}
\end{center}
\end{table}
\def\arraystretch{1}  


\section{Conclusions \label{sec:conclusions}}

We have presented a very simple way of calculating Coulomb and form-factor corrections to the $pp$ scattering amplitude in the context of an eikonal model. As we have emphasized, our approach is much simpler than that of Cahn \cite{Cahn} and  Kundr\'{a}t and Lokaji\v{c}ek  \cite{KL-Coulomb} which seem to have become standard in the analysis of Coulomb-nuclear interference effects at high energies. It is essentially model independent, with the corrections holding for any reasonable eikonal model which fits the total $pp$ scattering cross section and the forward slope parameter $B$ and gives a reasonable description of $d\sigma/dq^2$.

We have given parametrizations of corrections that hold at least from 100 GeV to 20 TeV, and illustrated the magnitude of the effects in the effective real part of the nuclear amplitude which interferes with the Coulomb amplitude in an appropriate phase convention. We have used the resuts to investigate the constant-phase approximation used in recent analyses of very-high-energy scattering, and introduced a very simple model for the correct $q^2$ dependence of the phase based on the location of the zeros in the real and imaginary parts of the nuclear component of the amplitude.

\begin{acknowledgments}
L.D.  would  like to thank the Aspen Center for Physics for its hospitality and for its partial support of this work under NSF Grant No. 1066293.  P.H.\ would like to thank Towson University Fisher College of Science and Mathematics for support.
\end{acknowledgments}

\appendix*

\section{The effective spin-averaged eikonal amplitude \label{sec:appendix}}

It is generally assumed that spin effects will be very small at very high energies, with the scattering dominated by absorptive effects that are insensitive to spin. However, a complete description of the $pp$ scattering matrix involves a number of spin-dependent amplitudes. These are conveniently labeled by the helicities of the incident and final protons following Jacob and Wick \cite{JacobWick}. For a process $a+b\rightarrow c+d$ where the initial and final particles have helicities $\lambda_a,\,\lambda_b,\,\lambda_c,\, \lambda_d$, the scattering amplitude $f_{\lambda_c,\lambda_d;\lambda_a,\lambda_b}$ assumes the form
\be
\label{Tmatrix}
f_{\lambda_c,\lambda_d;\lambda_a,\lambda_b}(s,q^2) = \frac{i}{2p}\sum_j(2j+1)\left(\delta_{\lambda_a\lambda_c}\delta_{\lambda_b\lambda_d}-S_{\lambda_c,\lambda_d;\lambda_a,\lambda_b}\right)d^j_{\lambda_a-\lambda_b;\lambda_c-\lambda_d}(\cos{\theta}),
\ee
where $j$ is the total angular momentum and the functions $d^j_{\lambda_a-\lambda_b;\lambda_c-\lambda_d}(\cos{\theta})$ are the standard rotation coefficients in the convention of Rose \cite{Rose}. The differential cross sections for specific helicity states are
\be
\label{sigma_spin}
d\sigma_{\lambda_c,\lambda_d;\lambda_a,\lambda_b}\big/dq^2 = \frac{\pi}{p^2}\lvert f_{\lambda_c,\lambda_d;\lambda_a,\lambda_b}(s,q^2)\rvert^2,
\ee
while the spin-averaged differential cross section is
\be
\label{sigma_av}
\frac{d\sigma}{dq^2} = \frac{\pi}{p^2} \frac{1}{(2s_a+1)(2s_b+1)}\sum_{\lambda_c,\lambda_d,\lambda_a,\lambda_b} \lvert f_{\lambda_c,\lambda_d;\lambda_a,\lambda_b}(s,q^2)\rvert^2.
\ee

The number of independent $S$-matrix elements for total angular momentum $j$ is restricted by time reversal $(S^j_{\lambda_a,\lambda_b:\lambda_c,\lambda_d}=S^j_{\lambda_c,\lambda_d;\lambda_a,\lambda_b})$, parity $(S^j_{-\lambda_c,-\lambda_b;-\lambda_a,-\lambda_b}=S^j_{\lambda_c,\lambda_d;\lambda_a,\lambda_b})$, and, for $pp$ scattering, the identity of the particles, $(S^j_{\lambda_a,\lambda_b;\lambda_c,\lambda_d}=S^j_{\lambda_c,\lambda_d;\lambda_a,\lambda_b})$, leaving five independent elements. These are conventionally taken, following \cite[Sec.\ IV]{GGMW}, as 
\be
\label{ind_amps}
S^j_{\frac{1}{2},\frac{1}{2};\frac{1}{2},\frac{1}{2}},\ S^j_{\frac{1}{2},-\frac{1}{2};\frac{1}{2},-\frac{1}{2}},\ S^j_{\frac{1}{2},\frac{1}{2};-\frac{1}{2},-\frac{1}{2}},\ S^j_{\frac{1}{2},-\frac{1}{2};-\frac{1}{2},\frac{1}{2}},\ {\rm and\ } S^j_{\frac{1}{2},\frac{1}{2};\frac{1}{2}-\frac{1}{2}}.
\ee

The first two, and the equal corresponding diagonal amplitudes $S^j_{-\frac{1}{2},-\frac{1}{2};-\frac{1}{2},-\frac{1}{2}},\ S^j_{-\frac{1}{2},\frac{1}{2};-\frac{1}{2},\frac{1}{2}}$, involve no helicity flips and are expected to be dominated at high energies by diffractive scattering, with no significant dependence on the helicities involved. Under this condition, those $S$-matrix elements and the corresponding scattering amplitudes are all approximately equal. 

The independent diagonal scattering amplitudes are
\ba
\label{hel_amp1}
f_{\frac{1}{2},\frac{1}{2};\frac{1}{2},\frac{1}{2}}(s,q^2) &=& \frac{i}{2p}\sum_j(2j+1)\left(1-S^j_{\frac{1}{2},\frac{1}{2};\frac{1}{2},\frac{1}{2}}\right)d^j_{0,0}(\cos{\theta}) \\
\label{hel_amp2}
&\approx& i\int_0^\infty db\,b\left(1-S_{\frac{1}{2},\frac{1}{2};\frac{1}{2},\frac{1}{2}}(s,b)\right)J_0(qb), \\
\label{hel_amp3}
f_{\frac{1}{2},-\frac{1}{2};\frac{1}{2},-\frac{1}{2}}(s,q^2) &=& \frac{i}{2p}\sum_j(2j+1)\left(1-S^j_{\frac{1}{2},-\frac{1}{2};\frac{1}{2},-\frac{1}{2}}\right)d^j_{1,1}(\cos{\theta}) \\
\label{hel_amp4}
&\approx& i\int_0^\infty db\,b\left(1-S_{\frac{1}{2},-\frac{1}{2};\frac{1}{2},-\frac{1}{2}}(s,b)\right)J_0(qb).
\ea

In these expressions, we have converted the sums over $j$ to integrals over the impact parameter $b=\sqrt{j(j+1)}\big/p$, and used  asymptotic relations between the rotation coefficients and Bessel functions derivable from known results on the relation between Jacobi polynomials and Bessel functions for $j$ large,  \cite[Sec.\ 8.1]{szego}.  These relations are discussed in detail in \cite[Secs.\ IIIA, IIID]{LDlegendre}, where the rotation coefficients are expressed for large $j$ in terms of series of Bessel functions. 

In the limit of no significant helicity dependence, the $S$-matrix elements in Eqs.\ (\ref{hel_amp2}) and (\ref{hel_amp4}) have a common limit $S(b,s)$,
\be
\label{equal_amps}
S_{\frac{1}{2},\frac{1}{2};\frac{1}{2},\frac{1}{2}}(s,b) \approx S_{\frac{1}{2},-\frac{1}{2};\frac{1}{2},-\frac{1}{2}}(s,b)\approx S(s,b).
\ee
It follows that the diagonal scattering amplitudes can all be expressed in terms of a single helicity-independent amplitude $f(s,q^2)$, with
\be
\label{f_amp_defined}
f_{\frac{1}{2},\frac{1}{2};\frac{1}{2},\frac{1}{2}}(s,q^2) = f_{-\frac{1}{2},-\frac{1}{2};-\frac{1}{2},-\frac{1}{2}}(s,q^2) \approx f_{\frac{1}{2},-\frac{1}{2};\frac{1}{2},-\frac{1}{2}}(s,q^2) = f_{-\frac{1}{2},\frac{1}{2};\frac{1}{2},-\frac{1}{2}}(s,q^2) \approx f(s,q^2), 
\ee
where
\be
\label{hel_amp5}
f(s,q^2) =  i\int_0^\infty db\,b\left(1-S(s,b)\right)J_0(qb).
\ee

Neglecting the presumably very small helicity-flip amplitudes, the spin-averaged differential cross section in \eq{sigma_av} becomes simply
\be
\label{sigma_av_2}
\frac{d\sigma}{dq^2} \approx \pi\lvert f(s,q^2) \rvert^2.
\ee
This was the the form used in the text, and used without discussion in most treatments of Coulomb-nuclear interference at high energies. In the limit described, it is not necessary to distinguish the two independent diagonal amplitudes, and the off-diagonal elements of the scattering matrix become relevant only when polarization and spin-correlation phenomena,  presumably small, are studied, as in \cite{Buttimore}.

In the absence of Coulomb effects, the total nuclear cross sections for specific initial helicity states are related through standard unitarity arguments and the optical theorem to the imaginary parts of the corresponding diagonal amplitudes $f_{\lambda_a,\lambda_b;\lambda_a,\lambda_b}(s,q^2)$ at $q^2=0$,
\be
\label{sigma_tot_hel}
\sigma_{tot;\lambda_a,\lambda_b}(s) = 4\pi \Im f_{\lambda_a,\lambda_b;\lambda_a,\lambda_b}(s,0).
\ee
Since, in the limit discussed above, these amplitudes have the common value $f(s,q^2)$, the total cross section for the nuclear scattering averaged over the initial helicities is just
\be
\label{sig_tot_av}
\sigma_{tot} = 4\pi \Im f_N(s,0),
\ee
 where we have included the label $N$. This is usual connection.
 
As emphasized Buttimore, Gotsman, and Leader in \cite{Buttimore}, double nuclear helicity-flip amplitudes can interfere with the corresponding magnetic interactions between the protons, and, if large enough, can potentially disrupt the extraction of the $\rho$ parameter from interference effects in the small-$q^2$ scattering region when $\rho$ is small. It is therefore useful to estimate the size of these helicity-dependent amplitudes. 

In the high-energy regime with which we are concerned, the leading contributions to those amplitudes are expected to arise mainly from the Regge exchange amplitudes associated with the $\rho$ and $\omega$ trajectories. The presence of these exchanges in $pp$ scattering is well-established, and the magnitude of the exchange amplitudes can be determined from the total cross sections. It is known, in particular, that the amplitudes decrease in magnitude with increasing energy as $s^{\alpha(0)-1}$ with $\alpha(0) \approx 1/2$ the Regge intercept at $q^2=0$.

The complete helicity-dependent Regge amplitudes for these exchanges can be constructed using the methods of King, Durand, and Wali \cite{KingDurandWali}. The helicity structure is essentially determined by the leading physical resonance associated with the trajectory, that is, the vector $\rho$ and $\omega$ mesons in the present case. These amplitudes therefore have the same structure as the electromagnetic interactions \cite{VertexFuncs}, and can interfere with them.  

From the results on the exchange amplitudes in \cite{eikonal2015}, we find that the real parts of those amplitudes at $q^2=0$ are about 70\% of the real part of the spin-independent amplitude at $W=50$ GeV; this ratio decreases to 8\% at 100 GeV, and 0.3\% at 1 TeV.  We expect the magnetic parts of the exchange amplitudes to have similar magnitudes.  The expected interference effects between the exchange amplitudes and the electromagnetic magnetic-moment amplitude should be reduced by similar factors relative to  the Coulomb-nuclear interference term in the charge sector. The contributions to the cross sections are further reduced at small $q^2$ by the factor $q^2/4m^2$ familiar for pure magnetic moment scattering. We conclude that these effects are negligible at high energies and that the potential problem noted in \cite{Buttimore} does not actually exist at these energies. The magnetic scattering can therefore be neglected, as in the main text. 
 
 The estimated contribution of the helicity-dependent exchange amplitudes themselves to the cross section through either single- or double helicity-flip terms is also small, on the order of 4\% at 10 GeV, decrease rapidly with increasing energy, and can also be neglected.

\bibliography{CoulombMathbib}

\end{document}